\documentclass{article}
\usepackage{graphicx}
\usepackage{multirow}
\usepackage{float} 
\usepackage{url,tabularx,amsfonts,slashed,amsmath,array}

\begin{document}
\begin{flushright}
preprint SHEP-12-40\\
preprint FR-PHENO-2012-037\\
\today
\end{flushright}
\vspace*{1.0truecm}
\begin{center}
{\large\bf The role of charge and spin asymmetries in profiling $Z'\to t\bar t$ events}\\
\vspace*{1.0truecm}
{\large L. Basso$^1$, K. Mimasu$^2$ and S. Moretti$^{2,3}$}\\
\vspace*{0.5truecm}
{\it $^1$Physikalisches Institut, Albert-Ludwigs-Universit\"at Freiburg\\
D-79104 Freiburg, Germany}\\
\vspace*{0.25truecm}
{\it $^2$School of Physics \& Astronomy, University of Southampton, \\
Highfield, Southampton, SO17 1BJ, UK}\\
\vspace*{0.25truecm}
{\it $^3$Particle Physics Department\\ Rutherford Appleton Laboratory\\
Chilton, Didcot, Oxon OX11 0QX, UK}\\
\end{center}

	\vspace*{0.5truecm}
	\begin{center}
\begin{abstract}
We study the sensitivity of top pair production at the Large Hadron Collider (LHC) 
to the nature of an underlying $Z'$ boson, including full tree level standard model background 
effects and interferences while assuming realistic final state reconstruction efficiencies. 
We demonstrate that exploiting asymmetry observables represents a promising way to distinguish 
between a selection of benchmark $Z'$ models due to their unique dependences on the 
chiral couplings of the new gauge boson.
\end{abstract}
\end{center}
	\section{Introduction}
	$Z'$ bosons are a ubiquitous feature of theories beyond the Standard Model (SM) arising from various BSM scenarios 
	such as $U(1)$ gauge extensions of the SM motivated by supersymmetry 
	or grand unified theories, Kaluza-Klein excitations of SM gauge fields or excitations of 
	composite exotic vector mesons in technicolor theories to name a few.

	Typically, such resonances are searched for at hadron colliders via the Drell-Yan (DY) 
	production of a lepton pair, i.e., $pp(\bar p) \to (\gamma,Z,Z') \to \ell^+\ell^-$, 
	where $\ell=e,\mu$. The Tevatron places limits on the $Z'$ mass, $M_{Z'}$, at around 1 TeV~\cite{Tevatron} 
	(for a sequential $Z'$) while the latest LHC limits lie around 2.3 TeV~\cite{LHC} from this channel. 
	Several phenomenological studies on how to measure the $Z'$ properties and couplings to SM 
	particles in this clean DY channel have been performed.

	These proceedings summarise a recently published paper~\cite{Basso:2012sz} addressing 
	the use of the top-antitop final state, i.e., $pp(\bar p)\to (\gamma,Z,Z') \to t\bar t$, 
	to probe these $Z'$ properties. While it may not have as much `discovery' scope as the DY channel, 
	owing to the large QCD background combined with the complex six-body final state and the 
	associated poor reconstruction efficiency, it remains important to extract the couplings 
	of new physics to the top quark. Furthermore, the fact that the top decays before hadronising, 
	transmitting spin information to its decay products, allows for the definition of spin asymmetry 
	observables which provide an extra handle on $Z'$ couplings not present in non-decaying final states. 

	We study the scope of the LHC to profile a $Z'$ boson mediating $t\bar t$ production, in both 
	standard kinematic variables as well as spatial/spin asymmetries, by adopting some benchmark 
	scenarios for several realisations of the sequential, Left-Right symmetric and $E_6$ based 
	$Z'$ models (specifically, the same as those in~\cite{Accomando:2010fz}). 
	Specifically, the issue of distinguishability of various models using these observables is addressed.
	
	\section{Asymmetries and $Z'$ couplings}\label{sec:defcalc}
	We define the asymmetry observables considered with the aim of determining their power to discriminate between $Z'$s. 
	We refer the reader to our paper for a more detailed discussion on these as well as the selection of benchmark models, 
	statistical uncertainties and definitions of significance. This study investigated charge (spatial) and spin 
	asymmetries and their dependence on top couplings to profile and distinguish the models considered. 

	A selection of charge asymmetry variables were investigated with the most sensitive found to be $A_{RFB}$, 
	defined by the rapidity difference of the top and antitop, $\Delta y=|y_t|-|y_{\bar{t}}|$, while also cutting 
	on the boost of the $t\bar{t}$ system. This increases the contribution from the $q\bar{q}$ initial state by 
	probing regions of higher partonic momentum fraction, $x$, where its parton luminosity is more important:
	\small
	\begin{align}
	    A_{RFB}&=\frac{N(\Delta y > 0)-N(\Delta y < 0)}{N(\Delta y > 0)+N(\Delta y < 0)}\Bigg |_{|y_{t\bar{t}}|>|y^{cut}_{t\bar{t}}}.
	\end{align}
	\normalsize
	The two spin asymmetries considered, termed double ($LL$) and single ($L$), are defined as follows:
	\small
	\begin{align}\label{asy_ALL}
	A_{LL}=\frac{N(+,+) + N(-,-) - N(+,-) - N(-,+)}{N_{Total}}\quad;\quad A_{L}=\frac{N(-,-) + N(-,+) - N(+,+) - N(+,-)}{N_{Total}}
	\end{align}
	\normalsize
	where $N$ denotes the number of observed events and its first(second) argument corresponds to the helicity of the top (anti)quark. 
	These observables are alternatively known as the spin correlation and spin polarisation asymmetries respectively and can be 
	extracted as coefficients in differential angular distributions of the top decay products.

	The dependence of the asymmetries on the vector and axial couplings $g_{V}$ and $g_{A}$ or alternatively, 
	the left and right handed couplings $g_{L}$ and $g_{R}$ of a generic neutral current are shown in~\cite{Basso:2012sz}.
	The charge asymmetry depends on the product of the vector and axial couplings of both the initial 
	and final state particle and can only be generated when all of these are non-vanishing.
	This is equivalent to measuring the relative magnitudes of their right and left handed couplings of the $Z^{\prime}$. For the spin asymmetries, $A_{LL}$ depends 
	on the couplings in a similar way to the total cross section, while $A_{L}$ is only non-vanishing for non-zero vector 
	and axial couplings of the final state tops. It is additionally sensitive to their relative sign i.e. it measures
	their relative handedness as with the charge asymmetry but only for the final state. 
	
	With these unique coupling dependences, it is our aim to show that asymmetries can provide extra information to distinguish $Z^{\prime}$ 
	models and ultimately contribute to extracting the couplings of an observed neutral resonance.
	\section{Results}\label{sec:results}
	We present a selection of results profiling the spatial and spin asymmetry distributions of the benchmark $Z^{\prime}$ 
	models compared to the SM including interference effects. The set of benchmarks are split into two categories: 
	those with a vanishing vector or axial coupling (the $E_{6}$ models with the `B-L' generalised left-right symmetric model) are 
	classed as the `$E_{6}$' type while the rest, with both couplings non-zero, are referred to as the `generalised' models. The variables 
	described in section~\ref{sec:defcalc} were computed as a function of the $t\bar{t}$ invariant mass within 
	$\Delta M_{t\bar t}=|M_{Z^{\prime}}-M_{t\overline{t}}|<500$ GeV and compared to the tree-level SM predictions. 
	The code exploited for our study is based on helicity amplitudes, defined through the HELAS subroutines~\cite{HELAS}, 
	and built up by means of MadGraph~\cite{MadGraph}. CTEQ6L1~\cite{cteq} Parton Distribution Functions (PDFs) were used, 
	with the factorisation/renormalisation scale at $Q=\mu=2m_t$. VEGAS~\cite{VEGAS} was used for numerical integration.
	\begin{figure}[h!]	
	\centering
	\includegraphics[width=0.32\linewidth]{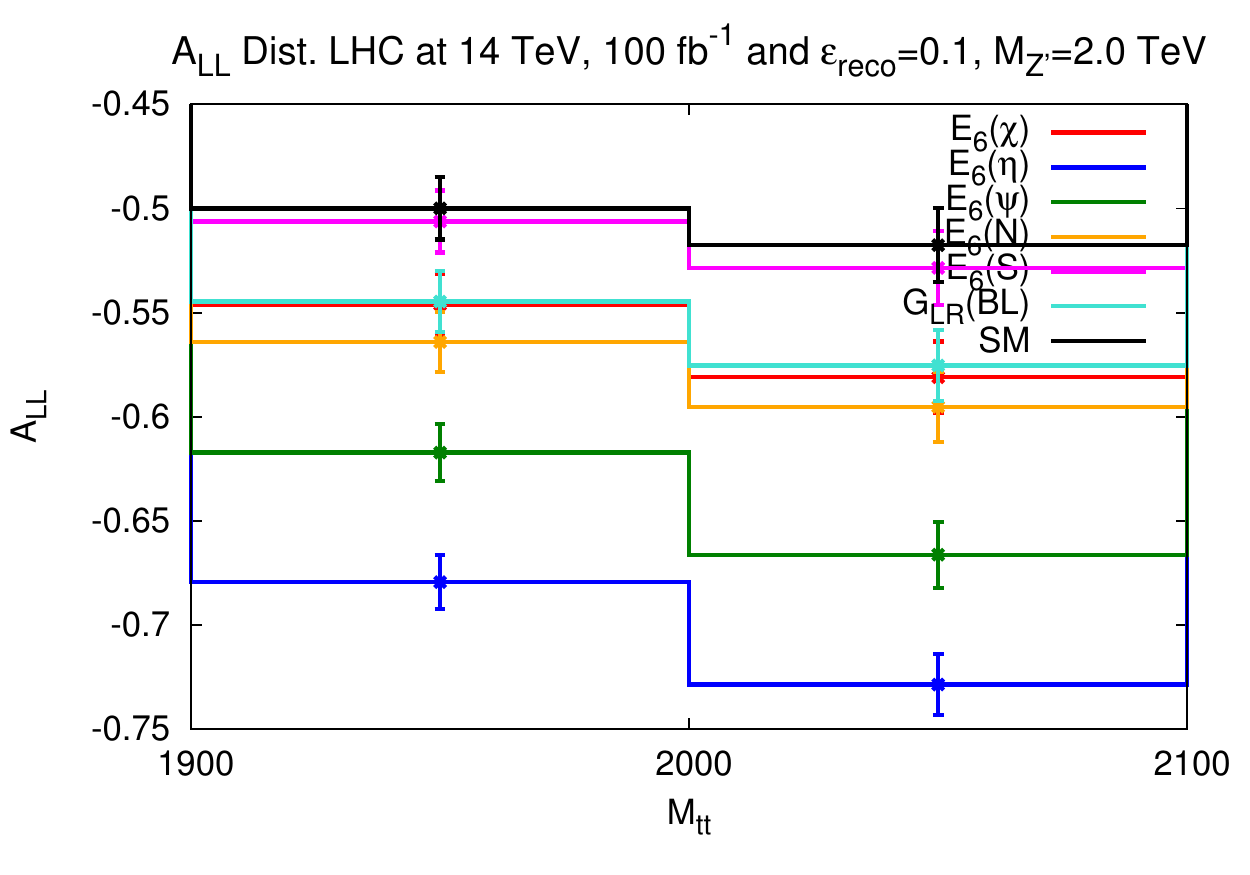}
	\includegraphics[width=0.32\linewidth]{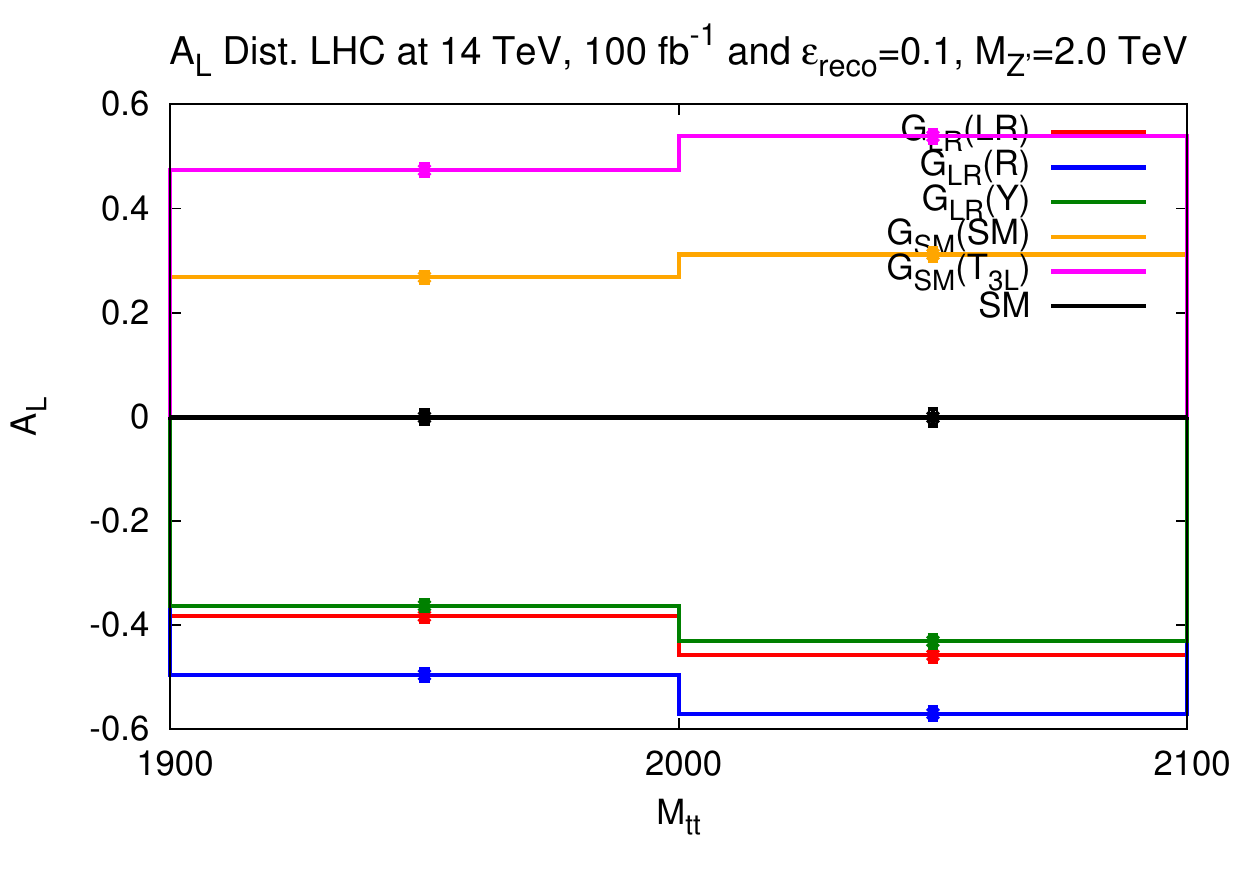}
	\includegraphics[width=0.32\linewidth]{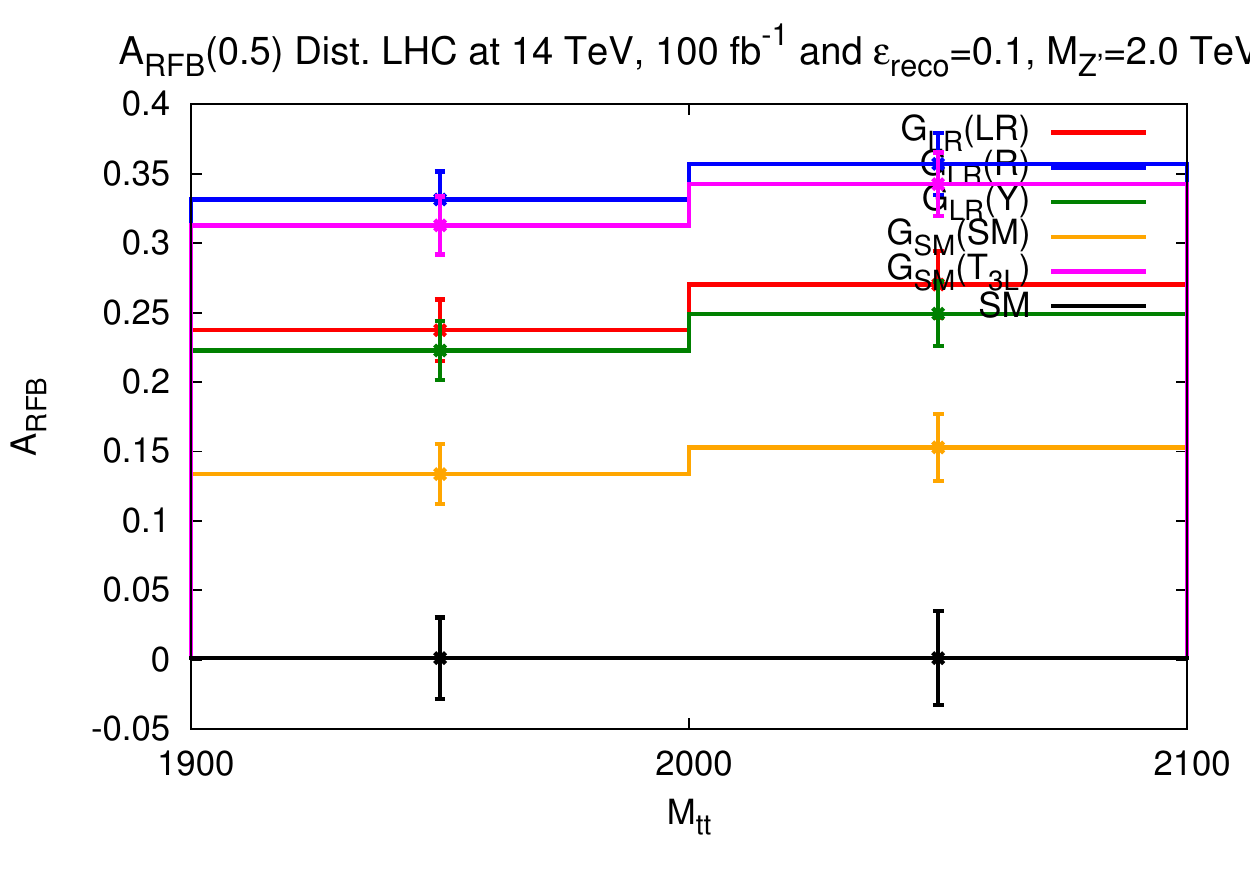}
		\caption{$A_{LL}$ for the $E_{6}$-type models and $A_{L}$ and $A_{RFB}(|y_{t\bar{t}}|>0.5)$ for the generalised models
	binned in $M_{t\bar{t}}$ 100 GeV either side of $M_{Z'}=$2 TeV for the LHC at 14 TeV assuming 
	100 fb$^{-1}$ of integrated luminosity.}\label{fig:asys}
	\end{figure}
	
	The asymmetry profiles shown in figure~\ref{fig:asys} include statistical uncertainties and fold in an estimated 10\% reconstruction efficiency of the 
	$t\bar{t}$ pair assuming the use of all possible decay channels. The figures show that the majority of benchmark models can be 
	distinguished from one another using these variables, noting in particular the sensitivity of 
	$A_{L}$ to the relative sign of the vector and axial couplings which allows for a clear distinction between 
	the $G_{SM}$ (sequential) and $G_{LR}$ (left-right symmetric).
	 
	$A_{LL}$ depends on the couplings in the same way as the total cross section and therefore models that cannot 
	be distinguished in the invariant mass spectrum will remain so in this observable. It is clear that $A_{L}$ is the most powerful observable
	in that it provides the best distinguishing power along with the extra feature of being sensitive to the handedness 
	of the top couplings. 
	$A_{RFB}$ provides some distinction but the handedness sensitivity is not present for reasons discussed in the previous section. Increasing the $Z^\prime$ mass 
	increases the statistical uncertainties but also slightly raises the central values, both as a consequence of the lower SM background. 
	Table~\ref{tab:signif_LHC14_AL} shows the significance of $A_{L}$ between benchmarks assuming 100 fb$^{-1}$ of 
	integrated luminosity. In almost all cases it is effective in disentangling the `generalised' models with the discrimination 
	decreasing slightly for higher masses. Although not shown in these proceedings, a similar statistical analysis was performed in determining how much integrated luminosity 
	would be required to achieve a significance of 3$\sigma$ between models using asymmetries. We showed that in 
	most cases, the models can be distinguished at the relatively early stages of the LHC ($\sim$~100~fb$^{-1}$ at 
	14 TeV) even for the higher mass of 2.5 TeV. The cases where this is not possible reflect mostly instances where the couplings are too similar.
	\begin{table}[h!]
	\centering
	\small
		\begin{tabular}{|c|c|c|c|c|c|c|}
			\hline
		$A_L$        &$SM$&$G{LR}(LR)$ &$G{LR}(R)$&$G{LR}(Y)$&$G{SM}(SM)$&$G{SM}(T_{3L})$\\ 
		\hline
		$SM$         & --         & 31.9(11.1) & 40.6(18.3) & 30.1(11.2) & 22.1(9.8)  & 38.7(22.5)\\
		$G{LR}(LR)$ & 16.9(7.7)  & --         & 10.0(6.6)  & 2.0(0.1)   & 62.2(21.7)  & 81.3(34.5) \\
		$G{LR}(R)$  & 21.3(11.5) & 4.6(4.0)   & --         & 12.0(6.5)  & 72.2(30.4) & 91.3(44.1) \\
		$G{LR}(Y)$  & 16.3(7.8)  & 1.0(0.1)   & 5.8(3.9)  & --         & 60.2(21.8)  & 79.3(34.6) \\
		$G{SM}(SM)$ & 11.8(6.3)  & 33.1(14.8) & 38.8(18.8) & 33.0(14.9) & --          & 19.1(13.7) \\
	$G{SM}(T_{3L})$& 20.1(13.9) & 42.5(23.0) & 48.5(27.2) & 42.7(23.2) & 9.7(7.8)  &  --    \\
		\hline
		\end{tabular}\caption{Significance for $A_{L}$ values around the $Z^{\prime}$ peak of generalised models, for the LHC at 14 
		TeV only. Upper triangle for $M_{Z'}=2.0$ TeV and lower triangle for $M_{Z^{\prime}}$=2.5 TeV. Figures refer 
		to $\Delta M_{t\bar t}<100$(500) GeV.}\label{tab:signif_LHC14_AL}
		\normalsize
		\end{table}
	\section{Conclusion}\label{sec:summary}
	We have presented an overview of a phenomenological study on classes of $Z^{\prime}$ models in both spin and 
	spatial asymmetries of $t\bar{t}$ production and showed that there is much scope to observe deviations from 
	the SM and even distinguish between various models, particularly for spin asymmetries. This suggests that the 
	$t\bar{t}$ channel may be a useful complement to the DY channel in the aim of profiling 
	a $Z'$ resonance should one be observed in the near future.

	We note that the benchmarks considered are put forward to set bounds on $Z^{\prime}$ masses and are 
	best probed in the di-lepton channels. Other models could be better suited to the $t\bar{t}$ channel, such as 
	leptophobic/top-phillic $Z^{\prime}$s occurring in composite/multi-site and extra-dimensional models. The profiling 
	techniques discussed in this study would be even more applicable in these scenarios.
	
	Finally, although not addressed in the paper to which these proceedings refer, one can ask whether more can be done 
	in profiling an observed resonance with the view of extracting its fermionic couplings. A previous study attempting to 
	do this in the light lepton sector~\cite{Petriello:2008zr} finds a degeneracy in determining the quark and lepton couplings 
	which can be solved by considering asymmetries in an alternate final state. In a more recent paper~\cite{Basso:2012ux}, we show 
	that asymmetry observables in the $t\bar{t}$ provide independent information which would break this degeneracy and allow 
	for a fit to all couplings in the `minimal' framework of 5 independent couplings used in many benchmarks.
	\section*{Acknowledgments}
	The work of KM and SM is partially supported through the NExT Institute. LB is supported by the Deutsche Forschungsgemeinschaft 
	through the Research Training Group grant GRK\,1102 \textit{Physics of Hadron Accelerators}. We would like to thank E. Alvarez 
	for pointing out the discussion on systematic uncertainties. KM would additionally like to thank the organisers of TOP2012 for their financial support.

\end{document}